# Current-induced energy barrier suppression for electromigration from first principles

Ruoxing Zhang[1,2], Ivan Rungger[2], Stefano Sanvito[2]*, Shimin Hou[1]*

[1] Key Laboratory for the Physics and Chemistry of Nanodevices, Department of Electronics, Peking University, Beijing 100871, China
[2] School of Physics and CRANN, Trinity College, Dublin 2, Ireland

**Abstract**

We present an efficient method for evaluating current-induced forces in nanoscale junctions, which naturally integrates into the non-equilibrium Green's function formalism implemented within density functional theory. This allows us to perform dynamical atomic relaxation in the presence of an electric current while also evaluating the current-voltage characteristics. The central idea consists in expressing the system energy density matrix in terms of Green's functions. In order to validate our implementation we perform a series of benchmark calculations, both at zero and finite bias. Firstly we evaluate the current-induced forces acting over an Al nanowire and compare them with previously published results for fixed geometries. Then we perform structural relaxation of the same wires under bias and determine the critical voltage at which they break. We find that, while a perfectly straight wire does not break at any of the voltages considered, a zigzag wire is more fragile and snaps at 1.4 V, with the Al atoms moving against the electron flow. The critical current density for the rupture is estimated to be $9.6\times10^{10}$A/cm$^2$, in good agreement with the experimentally measured value of $5\times10^{10}$A/cm$^2$. Finally we demonstrate the capability of our scheme to tackle the electromigration problem by studying the current-induced motion of a single Si atom covalently attached to the sidewall of a (4,4) armchair single-walled carbon nanotube. Our calculations indicate that if Si is attached along the current path, then current-induced forces can induce migration. In contrast, if the bonding site is away from the current path, then the adatom will remain stable regardless of the voltage. An analysis based on decomposing the total force into a wind and an electrostatic component, as well as on a detailed evaluation of the bond currents, shows that this remarkable electromigration phenomenon is due solely to the position-dependent wind force.

## I. INTRODUCTION

In recent years nanoscale devices have attracted increasingly large attention. The interest is motivated by their potential as a viable technology for either extending or replacing the conventional Si MOSFET platform.[1,2] However, as the device size shrinks, atomic rearrangements[3-7] and diffusion of atoms[8-12] in the presence of electrical currents become key problems, limiting the device mechanical stability and the persistence in time of uniform electron transport properties. These issues are closely related to the presence of current-generated forces acting upon the nuclei. The interaction between current-carrying electrons and ions manifests itself in two different ways, namely as local ionic heating and as current-induced forces.[13-15] Certainly they both may have significant effects on the atomic and electronic structure of a nanoscale junction.[13-23] Local heating involves a series of inelastic transitions among states of different energy, and thus it is associated with the excitations of the corresponding vibrational modes. Current-induced forces in contrast mainly arise from the charge density redistribution caused by the electron flow. In a pictorial way this force is analogous to the one exerted by the running water of a river upon the stones in its bank, and is usually referred to as the wind-force.

In the quasi-ballistic transport regime, the one investigated here, local ionic heating is small and usually gives an insignificant contribution to the atomic motion, but what about current-induced forces? Interestingly, the typical current densities in nanoscale junctions are much larger than those in the conducting interconnects widely used in solid-state circuits. Since current-induced electromigration is already one of the major causes of device failure in microelectronics,[24-27] we expect that in nanodevices current-induced forces will play an even more important role in limiting their strutural stability. This clearly indicates that a deep understanding of current-induced forces may help us in the design of more robust devices possibly with longer lifetimes. Furthermore one may also speculate of using current-induced forces to operate a device, for instance as a tool for switching a resistor between different resistance values, or to assemble devices by drifting atoms at desired positions with electrical currents.

Despite their importance, to our knowledge, only a few fundamental studies[14,28] have focused on the implementation of current-induced forces in a practical algorithm. Furthermore much remains still to be understood about the electromigration process and the calculations of diffusion paths. With these two goals in mind we have developed an efficient computational scheme for the calculation of forces at finite bias and implemented it in the *ab initio* electronic transport code *Smeagol*[29-31]. *Smeagol* is based on the non-equilibrium Green's function (NEGF) method[32-34] and it is interfaced with the localized atomic orbital pseudopotential code *Siesta*[35,36], from which it obtains the density functional theory (DFT) Hamiltonian matrix. Our scheme enables us to perform structural relaxation at finite bias and at the same time to monitor the device current-voltage (*I-V*) characteristics. Furthermore, it potentially opens up the possibility of performing molecular dynamics simulations under electron current flow conditions.

The implementation of current-induced forces is a rather challenging task, since it involves the evaluation of the atomic forces in a non-equilibrium situation, where the total number of electrons in the simulation cell is not guaranteed and the total energy is not defined. For a closed, finite and time-independent system at equilibrium, the atomic forces are well defined and can be obtained from the conventional Hellmann-Feynman (HF) theorem[37]. In contrast, for an open and out of equilibrium situation we are no longer able to define the atomic forces from such a classical energy perspective and an alternative strategy is needed. To our knowledge there are two different starting points in the derivation of the atomic forces under the presence of a current. The first starts from a time-dependent Lagrangian mean-field theory[38,39], while the second is completely based on quantum-mechanical many-body dynamic theory[40]. By using this second strategy it is possible to define the atomic forces for a general quantum mechanical system as the time derivative of the expectation value of the ionic momentum operators.[15,28] The method holds true for the time-dependent situation as well. Although the physical origin and the formal definition of current-induced forces seem rather clear, in the past there has been some controversy over whether or not such forces are conservative[41]. More recently the controversy seems to have set pointing towards the non-conservative nature of current-induced forces.[7,42]

Our paper is organized as follows. In the next section we briefly introduce our computational methodology and the technical implementations adopted in *Smeagol*. Then we present a series of test cases for atomic forces calculated either at zero or at finite bias. In the case of finite bias calculations we choose a capacitor setup, where the electrodes are completely disconnected, so that no current flows. Such a setup allows us to verify the correctness of the NEGF-calculated electrostatic forces as these can be compared to those obtained by a corresponding DFT total energy calculation including a finite electric field[35,36]. We then present an investigation on the current-induced forces acting on Al point contacts, and compare the results to previous calculations for similar systems, where jellium leads were used. We also determine the critical bias leading to junction breaking for straight and zigzag wires, finding a rather good agreement with available experimental results. Finally, in the last section we present a series of numerical calculations demonstrating the capability of our implementation to tackle the electromigration phenomenon. Our test case is that of one Si adatom drifting along the path of the current in the vicinity of a (4,4) armchair single-walled carbon nanotube (SWCNT).

## II. METHODS

In order to define the atomic forces acting on a system sustaining a steady-state electric current we first provide a brief overview of the transformed HF theorem, which relates conservative atomic forces to classical energies. The HF theorem expresses the total force acting upon the $I$-th atom, $\vec{F}_I$, as the negative derivative of the total energy, $E$, with respect to its position, $\vec{R}_I$,

$$\begin{aligned}\vec{F}_I &= -\frac{\partial E(\vec{R}_I)}{\partial \vec{R}_I} \\ &= -\frac{\partial \langle \Psi | \hat{H}(\vec{R}_I) | \Psi \rangle}{\partial \vec{R}_I} \\ &= \left[ -\langle \Psi | \frac{\partial \hat{H}(\vec{R}_I)}{\partial \vec{R}_I} | \Psi \rangle \right] + \left[ -\langle \frac{\partial \Psi}{\partial \vec{R}_I} | \hat{H}(\vec{R}_I) | \Psi \rangle - \langle \Psi | \hat{H}(\vec{R}_I) | \frac{\partial \Psi}{\partial \vec{R}_I} \rangle \right]\end{aligned} \qquad (1)$$

Here, $\hat{H}(\vec{R}_I)$ is the many-electron Hamiltonian operator and $|\Psi\rangle$ is the associated many-electron normalized wave-function. The first term in the last equality of Eq. (1) is the well-known conventional HF force, while the second term is often referred to as the Pulay force[43]. This vanishes only if $|\Psi\rangle$ is an exact eigenstate of $\hat{H}$ or if the basis set does not depend parametrically on the ionic coordinates (as for a plan-wave basis set). In that particular case the Eq. (1) reduces to the conventional HF theorem.

In DFT the ground state total energy is a well-defined quantity. As such the atomic forces can be calculated by taking explicitly the derivative of the total energy with respect to the atomic positions as written in the first equality of Eq. (1). This is well documented and a detailed description of the implementation used in *Siesta* can be found in references [35, 36, 44]. The result can be generally written as the sum of two terms

$$\vec{F} = \vec{F}_{BS} + \vec{F}_C . \qquad (2)$$

Here $\vec{F}_{BS} = -\partial E_{BS} / \partial \vec{R}_I$ describes the force originating from the band structure (BS) contribution of the total DFT energy, $E_{BS}$, which is equal to the sum of the eigenvalues of the occupied states. The second term, $\vec{F}_C$, is obtained by taking the derivative of the remaining contributions to the DFT total energy. Importantly, the force given in Eq. (2) automatically includes also the Pulay corrections arising from the fact that the employed basis set is constructed with local atomic orbitals (numerical). The Kohn-Sham (KS) equation of the system reads

$$H_{\mu\nu}\psi_{i\nu} = \varepsilon_i S_{\mu\nu}\psi_{i\nu} , \qquad (3)$$

where $H_{\mu\nu}$ is the Kohn-Sham Hamiltonian matrix element, $S_{\mu\nu}$ is the overlap matrix element, $\varepsilon_i$ is the $i$-th KS eigenvalue, $\psi_{i\mu}$ is the corresponding eigenvector and the indices $\mu$ and $\nu$ label the local orbital basis set. The density matrix of the system, $\rho_{\mu\nu}$, is then defined as

$$\rho_{\mu\upsilon} = \sum_i f(\varepsilon_i)\psi_{i\mu}\psi_{i\upsilon}^* \ , \tag{4}$$

where $f(\varepsilon_i)$ is the occupation probability of the state having $\varepsilon_i$ as its KS eigenvalue. In our case $f(\varepsilon_i)$ is the Fermi-Dirac distribution. The band structure force, $\vec{F}_{BS}$, can then be written as

$$\vec{F}_{BS} = -\sum_{\mu\upsilon}\rho_{\mu\upsilon}\frac{\partial H_{\mu\upsilon}}{\partial \vec{R}_I} + \sum_{\mu\upsilon}\Omega_{\mu\upsilon}\frac{\partial S_{\mu\upsilon}}{\partial \vec{R}_I} \ , \tag{5}$$

where $\Omega_{\mu\upsilon}$ is the so-called energy density matrix, defined as

$$\Omega_{\mu\upsilon} = \sum_i \varepsilon_i f(\varepsilon_i)\psi_{i\mu}\psi_{i\upsilon}^* \ . \tag{6}$$

Since our interest is that of extending the formalism to open systems described by the NEGF formalism, it is convenient to re-write Eq. (4) and Eq. (6) in terms of the retarded Green's function $G_{\mu\upsilon}(E) = [(E+i\delta)S - H]^{-1}_{\mu\upsilon}$, with $\delta$ being a small positive number. The Eq. (4) can then be re-casted as energy integral

$$\rho_{\mu\upsilon} = -\frac{1}{2\pi i}\int_{-\infty}^{\infty}\left[G_{\mu\upsilon}(E) - G^{\dagger}_{\mu\upsilon}(E)\right]f(E)dE \ , \tag{7}$$

and it is straightforward to show that a similar expression holds also for the energy density matrix

$$\Omega_{\mu\upsilon} = -\frac{1}{2\pi i}\int_{-\infty}^{\infty}E\left(G_{\mu\upsilon}(E) - G^{\dagger}_{\mu\upsilon}(E)\right)f(E)dE \ . \tag{8}$$

The only difference between Eq. (7) and Eq. (8) is the additional factor $E$ appearing in the integrand. The advantage of expressing the forces in terms of Green's functions is that we can now extend the formalism to open systems.

We now move to define the forces for a system out of equilibrium and sustaining a steady-state current. In this situation the HF theorem is not directly applicable.[15,28,45,46] However, Di Ventra and co-workers have shown that the forces can be defined in a more general way, namely as the time derivative of the expectation value of the ionic momentum operator,[15,28] i.e. as

$$\vec{F}_I = \frac{\partial}{\partial t}\langle\Psi(t)|-i\hbar\frac{\partial}{\partial \vec{R}_I}|\Psi(t)\rangle \ , \tag{9}$$

where the wave function is in general time-dependent. Based on Eq. (9) and the more restrictive condition for the wave function[28]

$$\langle\Psi(t)|\hat{H} - i\hbar\frac{\partial}{\partial t}|\Psi(t)\rangle = 0 \ , \tag{10}$$

by some algebraic manipulations we eventually derive an Ehrenfest-like expression for the forces that applies to a generic time-dependent problem expanded over a finite basis set

$$\vec{F}_I = -\left[\left\langle \Psi(t)\left|\frac{\partial \hat{H}}{\partial \vec{R}_I}\right|\Psi(t)\right\rangle + \left\langle \frac{\partial \Psi(t)}{\partial \vec{R}_I}\left|\hat{H} - i\hbar\frac{\partial}{\partial t}\right|\Psi(t)\right\rangle + \left\langle \frac{\partial \Psi(t)}{\partial \vec{R}_I}\left|\hat{H} - i\hbar\frac{\partial}{\partial t}\right|\Psi(t)\right\rangle^*\right]. \qquad (11)$$

Note that for steady-state transport problems the wave function of the system can be written as $|\Psi(t)\rangle = e^{-iEt/\hbar}|\psi\rangle$, where E is a phase factor with units of energy. By using this result and Eq. (11), a final form for the total atomic forces in a system with steady-state electrical current flow is given as

$$\vec{F}_I = -\frac{\partial \langle \psi|\hat{H}|\psi\rangle}{\partial \vec{R}_I} \quad . \qquad (12)$$

At this point we want to stress that Eq. (12) is formally identical to the HF force given in Eq. (1), although the two equations are based on a completely different derivation, and Eq. (12) holds under more general conditions. Because of this formal analogy, however, the expression in Eq. (2), derived for a closed system at equilibrium, is also valid for the steady-state transport problem, which is explored here. Importantly, all the terms contained in $\vec{F}_C$ depend entirely on the charge density matrix $\rho_{\mu\nu}$ [35,44]. This means that the forces can be computed exactly as in a standard DFT ground state calculation once the non-equilibrium $\rho_{\mu\nu}$ is known. Therefore in the remaining part of this section we focus on detailing how $\rho_{\mu\nu}$ and $\Omega_{\mu\nu}$ are extracted in the non-equilibrium case.

The NEGF scheme implemented in *Smeagol* is based on dividing a two-terminal device into three parts: the semi-infinite left and right current/voltage electrodes (the leads) and the scattering region (or the extended molecule)[30]. The effects of the leads on the scattering region are taken into account via the energy-dependent self-energies, and the associated coupling matrices, $\Gamma_{L\mu\nu}(E)$ for the left electrode, and $\Gamma_{R\mu\nu}(E)$ for the right electrode. For such a system setup the non-equilibrium density matrix of the scattering region is given by

$$\rho_{M\mu\nu} = \frac{1}{2\pi i}\int_{-\infty}^{\infty} G^<_{M\mu\nu}(E)dE \quad , \qquad (13)$$

where the additional subscript $M$ indicates that the orbital indices $\mu,\nu$ run only over the basis set functions localized in the scattering region. $G^<_{M\mu\nu}(E)$ is the energy and voltage dependent lesser Green's function for the scattering region, and is given by

$$G^<_{M\mu\nu}(E) = \sum_{\alpha\beta} iG_{M\mu\alpha}(E)\left[f_L(E)\Gamma_{L\alpha\beta}(E) + f_R(E)\Gamma_{R\alpha\beta}(E)\right]G^\dagger_{M\beta\nu}(E). \qquad (14)$$

In the same way as for the equilibrium case [Eq. (8)], it is then straightforward to show that $\Omega_{\mu\nu}$ is calculated with an analogous equation to the one used to obtain $\rho_{\mu\nu}$ [i.e. Eq. (13)], but where the integrand is now multiplied by the energy $E$

$$\Omega_{M\mu\upsilon} = \frac{1}{2\pi i} \int_{-\infty}^{\infty} E G^{<}_{M\mu\upsilon}(E) dE. \tag{15}$$

Although both $\rho_{M\mu\upsilon}$ and $\Omega_{M\mu\upsilon}$ depend on the charge density of the entire infinite open system (extended molecule plus leads), they can be calculated by using just the Hamiltonian of the scattering region and the self-energies of the leads, which also set the appropriate open boundary conditions. At zero-bias Eq. (13) and Eq. (15) are equivalent to Eq. (7) and Eq. (8), once these are evaluated only over the basis indices running over the scattering region.

The scheme for calculating current-induced forces discussed up to this point has been implemented into the *Smeagol* code. Before going through a few examples, demonstrating our ability of performing structural relaxation at finite bias, we will now report some technical details of the implementation. In general, in order to ensure a more smooth convergence of the charge density, we always attach at each side of the scattering region one principal layer (unit cell) of the leads. Then, when performing structural relaxation, the atoms of such principal layers are always kept fixed at their equilibrium positions (the ones of the bulk crystal). We note that at zero-bias our formalism allows us to calculate the projection of the total energy onto the scattering region, just like in a standard ground state DFT calculation, as this ultimately depends only on $\rho_{M\mu\upsilon}$ and $H_{M\mu\upsilon}$. At finite bias a total energy is not defined, especially since the forces might not be conservative.[7,42]

Under current flow conditions the density matrix of the scattering region responds to the applied bias, producing a redistribution of the electron density at the surface of the electrodes and inside the atomic junction itself (typically a molecule). The charge density accumulation at the surface of the electrodes generates an electric field, similarly to what happens in a parallel plate capacitor. As a consequence there is a force of purely electrostatic nature (also denoted as "direct force"[47]), originating from the charge accumulation at the surface of the electrodes, acting on the ions located in between the two surfaces. In contrast, the local change in the charge density of the bridging molecule itself gives rise to a second contribution to the current-induced forces, which we refer to as wind force (the force described as originating from the continuum wave functions in Ref. [47]). In literature the term "wind force" often refers to the force originating purely from the momentum transfer from the electrons to the ions;[48,49] here we also include the forces originating from current-induced charge density rearrangements into the wind-force, since also these are ultimately caused by the electron flow (wind). For a given device it is then interesting to analyse the direct force and wind force independently.

Unfortunately purely electrostatic and wind forces are not observable separately, so that it is not formally possible to distinguish between the two. If we assume that there is a rather small bias induced charging of individual atoms in the bridging molecule, then we can use an approximate procedure[47,48]. Firstly, for fixed structure and fixed bias voltage the total force, $\vec{F}\left[\rho_{\mu\upsilon}(V);H_{\mu\upsilon}(V)\right]$, is calculated. Then an approximation for the electrostatic force can be obtained by calculating the forces for the equilibrium zero-bias charge density and the corresponding Hamiltonian, to which we add a potential shift, $\Delta H_{\mu\upsilon}$. Such a shift describes an electrostatic potential offset between the two electrodes equal to *V* and a linear potential drop within the scattering region (as expected for a parallel plate capacitor). Technically $\Delta H_{\mu\upsilon}$ is obtained by adding to the Hamiltonian of the left (right) lead the corresponding overlap matrix multiplied by $+eV/2$ ($-eV/2$) and a linear potential drop in the scattering region, i.e. $\Delta H_{\mu\upsilon}$ describes a position dependent shift of the matrix elements of the Hamiltonian of the scattering region due to the electric field. We can then define the electrostatic force, $\vec{F}_{Field}(V)$, as

$$\vec{F}_{Field}(V) = \vec{F}\left[\rho_{\mu\upsilon}(0);H_{\mu\upsilon}(0)+\Delta H_{\mu\upsilon}(V)\right]. \tag{16}$$

Then the wind force, $\vec{F}_{Wind}(V)$, is simply obtained by subtraction as

$$\vec{F}_{Wind}(V) = \vec{F}\left[\rho_{\mu\upsilon}(V);H_{\mu\upsilon}(V)\right] - \vec{F}_{Field}(V). \tag{17}$$

Clearly this definition is only operational and it is not applicable in general (for instance when the charge density is severely distorted by the bias). However by using this simple separation we can often provide a reasonable estimate of the two contributions and understand which force dominates in a particular device (see Sec. V).

## III.   FORCES FOR NON-CURRENT CARRYING SYSTEMS

As a first test for our methodology in this section we present a set of calculations for systems where there is no current flow. Our aim is to compare forces calculated with the NEGF scheme implemented in *Smeagol* (open boundary conditions) to those obtained from a DFT ground-state calculation using periodic boundary conditions (PBC), as performed with *Siesta*. In order to compare results at finite bias (but still with no current) we consider a capacitor setup, where the two electrodes are so well separated to be electronically non-interacting among each other.

### A.   Atomic forces at equilibrium: one-dimensional Au monatomic chain

Our first goal is that of verifying the validity of the Green's function approach in calculating the energy density matrix at zero-bias [Eq. (8)] in a practical situation. To this goal we choose a simple and idealized one-dimensional (1D) system, where the scattering region consists of a gold chain of 9 atoms, connected to 1D gold electrodes. The lattice spacing is assumed to be uniform, 2.8 Å, with the leads unit cell containing two atoms. We then shift the middle Au atom in the scattering region by 1 Å in the direction transverse to the chain [see Fig. 1(a)], so that a rather large restoring force is expected. In Fig. 1(b) we present the calculated forces, where the solid black lines represent those obtained with a standard *Siesta* calculation for a system periodic along the *z* direction, whereas the dashed red lines represent the forces obtained by using our NEGF scheme. Note that all the forces are distributed in such a way to move the shifted atom back into the chain and that there is essentially no difference between the two methods. This confirms that the NEGF-calculated charge density is identical to numerical precision to that calculated with standard DFT and PBC, and also that the equilibrium $\Omega_{M\mu\upsilon}$ is calculated correctly by Eq. (15). In order to emphasize further the importance of calculating correctly $\Omega_{M\mu\upsilon}$, we also show the results for forces obtained by setting $\Omega_{M\mu\upsilon}=0$ (dash-dotted blue lines). These are considerably larger and for some atoms they even point in the wrong direction.

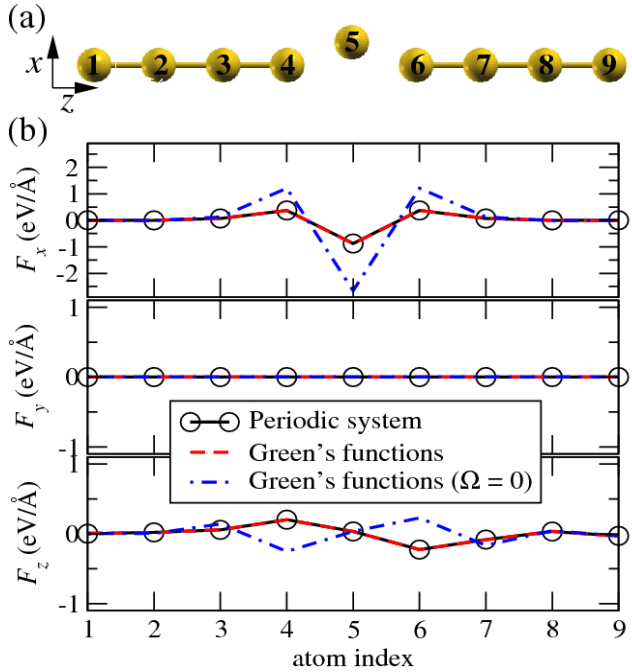

Fig. 1: (Color online) Zero-bias test for the atomic forces calculated from the NEGF scheme. The system investigated is a linear Au chain where one atom has been displaced from the chain axis (a). In (b) we show the *x*-, *y*- and *z*-component of the atomic forces acting on these nine Au atoms.

## B. Field induced forces over the surface atoms of the electrodes

In order to evaluate the accuracy of the calculated forces at finite bias, we now consider a parallel-plate capacitor that consists of two semi-infinite lithium electrodes separated by a 10 Å long vacuum gap [see Fig. 2(a)]. Each atomic layer contains 9 atoms, and we use 4×4 *k*-points in the *x-y* plane to account for periodic boundary conditions in the orthogonal direction. There are 4 atomic layers in each lead unit cell.

Fig 2(c) shows the planar average in the *x-y* plane of the difference between the electrostatic potential at finite and zero bias (dashed blue line) for a voltage of 2 V. Since the vacuum region electronically disconnects the two electrodes, the same calculation can be also performed as a ground state PBC calculation with an electric field applied along the *z*-axis (solid black line). Note that the discontinuity in the applied sawtooth-like potential is located in the middle of the vacuum region, and therefore does not affect the charge density. From the figure we note that at self-consistency the planar average of the potential is flat in the metal, while the field induced drop is located in the vacuum region. The resulting charge density for this system, and therefore also the forces, should be approximately equal across the two methods.

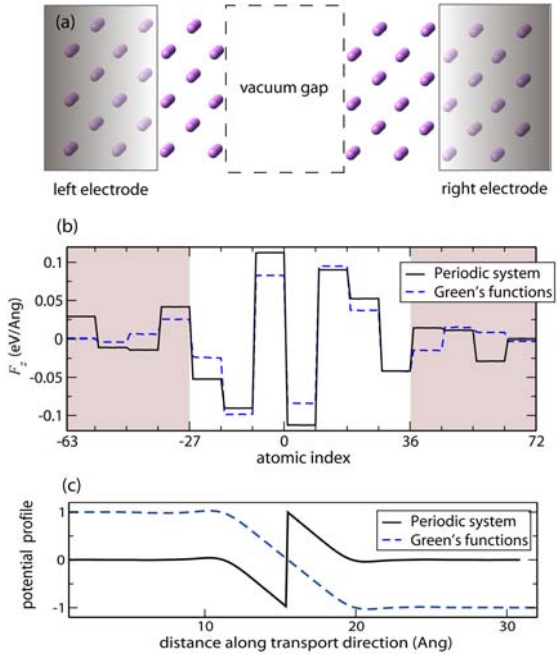

Fig. 2 (Color online) Zero-bias test for the forces acting on a parallel plate capacitor. (a) Schematic representation of the system investigated, namely a parallel-plate capacitor composed of two semi-infinite lithium bulk electrodes separated by 10 Å long vacuum gap. (b) Atomic forces acting of every atom calculated with both *Smeagol* and *Siesta* at zero-bias. (c) Planar average of the electrostatic potential for the system in (a) when a voltage of 2V is applied.

We start our analysis of the forces by discussing again the zero-bias case presented in Fig. 2(b). In the figure the atoms are sorted along the *z*-direction: a negative index indicates an atom located in the left electrode, while positive indexes are for those in the right electrode. Atom 0 is the right-most atom of the left electrode. We note that although both methods yield a similar trend, there are some non-negligible quantitative differences. Within the NEGF approach the forces acting on the atoms at the boundary of the scattering region are expected to be somewhat inaccurate, since it is the location where the scattering region is joined to the bulk semi-infinite leads.

However, this does not pose a problem since the boundary atoms belong to those lead principal layers which in *Smeagol* are always kept fixed at their bulk positions. More worryingly however is the fact that some forces are different at the electrode surface, where atomic relaxation should be performed. In order to investigate the origin of this discrepancy in more details, we have calculated the forces for the same system, but where we now add respectively 4 [red dashed lines in Figs. 3(a) and (b)] and 8 [black lines in Figs. 3(a) and (b)] additional Li atomic layers at each side of the scattering region. We note that the NEGF calculation returns us forces almost constant with the length. This reflects the fact that the present calculations effectively concern only two semi-infinite leads, so that the number of layers inside the scattering region should not affect the result. In contrast the forces calculated with PBC converge towards the NEGF result only as the electrodes get longer. This demonstrates that the forces are basically identical within two methods, as long as a sufficiently large number of atomic layers are included in the PBC calculation. It also shows that finite size effects are smaller in the NEGF approach than in the PBC calculations. As such we believe that our open boundary condition approach constitutes an important calculation platform for studying surface reconstruction, since it naturally includes the correct boundary conditions, in contrast to standard periodic calculations for a finite slab.

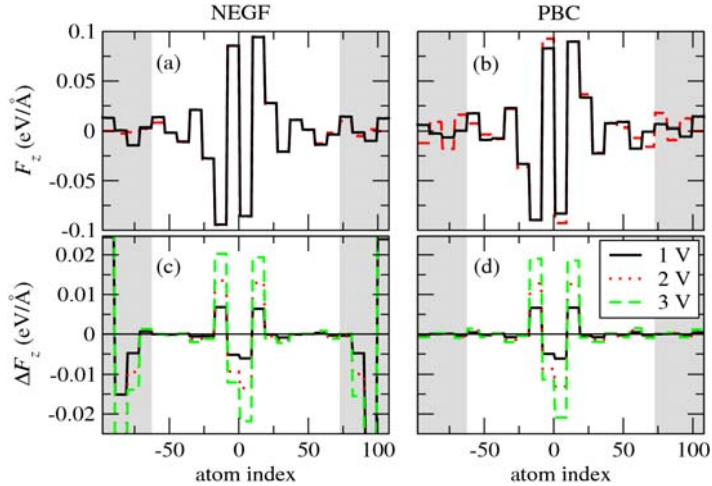

Fig. 3 (Color online) Finite-bias and zero-bias tests for the field-induced forces acting on a parallel plate Li capacitor as a function of the number of atomic layers included in the simulation cell. In (a) and (b) we show the forces at zero-bias for the system in Fig. 2(a), where 4 (dashed red lines) and 8 (solid black lines) additional Li layers have been added to each electrode; in (c) and (d) the bias induced change in the force is shown for different bias voltages (this is the difference between the forces at zero- and finite bias). Panels (a) and (c) show results obtained using the NEGF formalism, whereas the results in (b) and (d) are obtained by using PBC and an equivalent applied electric field.

We now move to analyze the forces calculated with the NEGF scheme at finite bias. These are presented in Fig. 3(c), where we display the difference between the forces at finite bias and those at zero-bias plotted as a function of the atomic position (in practice the atomic index). The same quantity is shown in Fig. 3(d) for a PBC calculation with an equivalent applied electric field. The most notable feature is the presence of spurious NEGF-calculated forces on the atoms at the boundary of the unit cell. This is due to the fact that at finite bias *Smeagol* introduces a potential step at this boundary, which results in a spurious contribution to the forces. However, again this only affects the peripheral atoms of the leads, i.e. those that are not relaxed. In the center of the scattering region the charge induced change in the forces is basically identical when calculated with the two methods. This confirms the correctness of Eq. (15). Note that the field induced forces at the electrode interface layers, which have been discussed in this section, are expected to produce the main contribution to the finite bias relaxation of the metal-insulator interface in tunnel junctions, where the currents are usually very small[50,51].

## IV. CURRENT-INDUCED FORCES IN Al NANOWIRES

So far we have only investigated systems in which there is no current flow. In this section we present results for current-induced forces and the associated bias-induced structural instabilities. In the last decade there have been a number of calculations on current-induced forces in metallic nanowires,[3,4,15,20,28,41,52,53] so that a few theory benchmarks exist. In order to test our implementation, we have decided to investigate the forces acting over a 4-atom long straight Al wire, a system, which was previously discussed by Di Ventra *et al.* in Ref. [20]. In Di Ventra's work the forces as function of bias were calculated for Al wires of different lengths connected to jellium leads. In our simulation we attach the 4-atom long Al chains to the hollow sites of flat Al(111) electrodes. These contain 9 Al atoms per plane and we include in the scattering region 5 atomic layers on each side of the wire [see Fig. 4(a)]. The equilibrium bond lengths are found by first relaxing the structure of a straight infinite Al monatomic wire, and then by minimizing the bond length between a finite 4-atom long chain and an Al slab. The second electrode is added in a perfectly symmetric way.

Once the cell is constructed we then relax further the 4-atom long wire, while keeping the electrodes atoms fixed. The final relaxed bond distances in the wire are respectively 2.46 Å, 2.45 Å and 2.46 Å, and the wire to surface distance is 2.02 Å. We note that in Ref. [20] the relaxed Al-Al bond length is uniform and equal to 3.069 Å. In all the calculations presented here we use a single-$\zeta$ plus polarization basis set, as control tests employing a basis of double-$\zeta$ plus polarization quality give essentially identical results (both at zero and finite bias). The real space mesh cutoff is equal to 400 Ry, and 4×4 k-points are used in the *x-y* plane. Note that in order to obtain correct results for the forces at finite bias under current flow conditions, it is not possible to avail of time-reversal symmetry in the *x-y* plane, so that one needs to sample the full Brillouin zone.

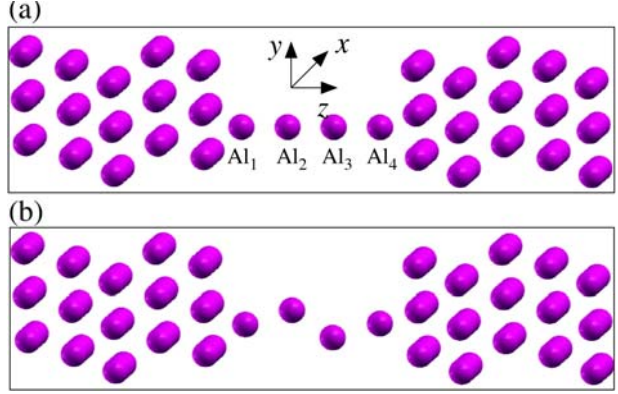

Fig. 4 (Color online) Schematic representations of the atomic structure of the straight (a) and the zigzag (b) 4-atom long Al wires anchored to Al electrodes investigated here.

In Fig. 5 we show the calculated forces as function of bias for the straight 4-atom long Al chains. These can be compared with those reported in Fig. 1(d) of Ref. [20] for bias values ranging between -1 V and 0 V (note that positive bias in Ref. [20] corresponds to negative bias in our calculations). In general we find a good agreement with the previously published results, except for the forces acting on atom 3 which are somehow different from those of Ref. [20]. The differences however are not significant and can be easily accounted for by the different leads used in the two calculations and by the consequent different initial bond lengths. In fact different atomic configurations result in slightly different charge distributions and these play an important role in determining the current-induced forces.

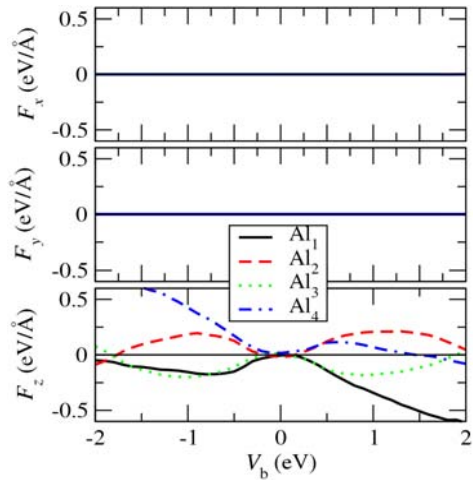

Fig. 5 (Color online) Forces as function of the applied bias voltage acting on a 4-atom long Al wire with straight configuration [see Fig. 4(a)]. Note that the forces are not zero only along the wire direction $z$.

We then perform structural relaxations of the atoms in the chain under bias, and find very little change in the atomic structure for bias voltages down to -3 V (we apply negative bias for the structural relaxations in order to be consistent with the bias direction used in Ref. [20]). A similar result was found previously for Au wires.[52] The reason for this seemingly very large stability of the chain structure under bias has to be found in the somehow artificial setup of the perfectly straight chain. This causes the forces along *x* and *y* to vanish almost exactly. In fact, in our calculations the *x* and *y* components of the forces are smaller than $10^{-3}$ eV/ Å at zero-bias and remain approximately constant for all voltages [figures 5(a) and (b)].

The forces vanish in the *x-y* plane due to an approximate rotational symmetry of the wire about the *z*-axis. In general symmetries relate the forces on the different atoms. If we denote the position of the midpoint of the bond between $Al_2$ and $Al_3$ as $x_0$, then the system is symmetric under reflection across the *x-y* plane passing for $x_0$. Therefore we have the following symmetry relations $F_x(V_b,Al_1)=F_x(-V_b,Al_4)$, $F_y(V_b,Al_1)= F_y(-V_b,Al_4)$ and $F_z(V_b,Al_1)=-F_z(-V_b,Al_4)$. The system is also approximately symmetric under rotations about the *x*-axis, with rotation center at $x_0$. This additional symmetry implies $F_x(V_b,Al_1)=F_x(-V_b,Al_4)$, $F_y(V_b,Al_1)=-F_y(-V_b,Al_4)$ and $F_z(V_b,Al_1)=-F_z(-V_b,Al_4)$. Mirror and rotation symmetry can only be fulfilled at the same time with $F_y(V_b,Al_1)= F_y(V_b,Al_4)=0$. If we also consider the rotation symmetry around the *y*-axis with rotation center at $x_0$, we obtain in an analogous way $F_x(V_b,Al_1)=F_x(V_b,Al_4)=0$. The same can be shown for atoms $Al_2$ and $Al_3$. From Fig. 5(c) we can see that for the remaining force along the *z*-direction we indeed have $F_z(V_b,Al_1)\approx-F_z(-V_b,Al_4)$ and $F_z(V_b,Al_2)\approx-F_z(-V_b,Al_3)$. We note again here that in order to obtain a zero force in the *x-y* plane for all bias voltages it is important to sample the *k*-points over the entire *x-y* Brillouin zone.

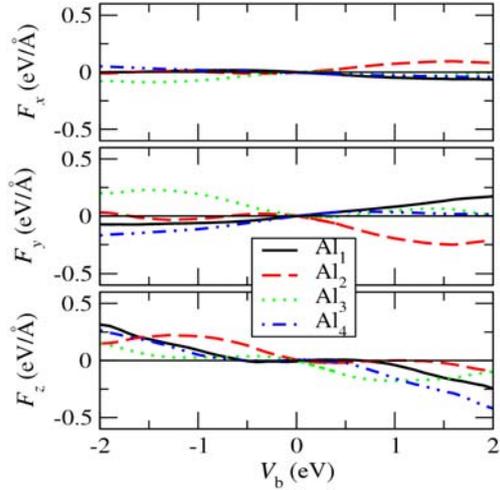

Fig. 6 (Color online) Forces as function of the applied bias voltage for a 4-atom long Al monatomic chain with zigzag configuration [see Fig. 4(b)]. Note that the forces are significant only in the zigzag plane (*y-z* plane).

In order to study the effects of the current-induced forces on a more realistic wire structure, we perform additional structural relaxation at zero-bias, this time by initializing the atomic coordinates of the two central atoms slightly off the wire axis. Interestingly the final relaxed structure presents a zigzag shape [Fig. 4(b)], with the zigzag plane mainly oriented along the *y*-axis (there is also a small shifts along the *x*-axis), so that the system is approximately symmetric under rotation about the x-axis at x0. This new structure has a total energy lower than that of the straight configuration, in agreement with previous studies[54]. In Fig. 6 the bias dependent forces acting on the atoms of the zigzag chain are shown. These are qualitatively different from those presented in Fig. 5 for the straight wire, since now there are large forces in all the directions. In particular the forces lie mainly in the plane spanned by the zigzag chain, while they are small along the direction perpendicular to the plane. This feature reflects well the approximate reflection symmetry across such a plane. We also note that the forces along the *z*-direction are substantially different from the ones calculated for the straight chain.

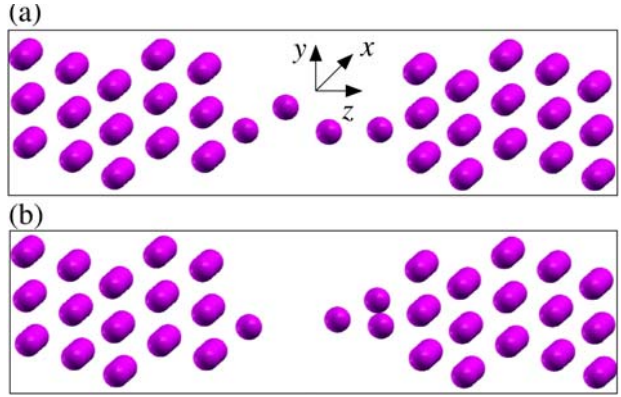

Fig. 7 (Color online) Relaxed structures at $V_b = -0.2$ V (a) and $V_b = -1.4$ V (b) obtained as the bias voltage increases from the original zigzag atomic configuration of Fig. 4(b).

We then perform structural relaxations under bias also for the zigzag setup and find dramatic changes in the structure as the voltage increases. Already for *V*=-0.2 V there is a transition to a different zigzag configuration, in which one atom only now lies off the *z*-axis [see Fig. 7(a)]. As the voltage is further increased we observe rather small but continuous changes in the structure. Finally at -1.4 V there is a new discontinuous structural change, which effectively corresponds to the wire breaking [see Fig. 7(b)]. This final structure does not get modified any longer by any voltage increasing down to -3 V. A wire breaking voltage of -1.4 V is in a rather good agreement with the experimentally found breaking voltages for Au and Ir nanowires [55]. For Al nanowires no such experimental data are available, however in Ref. [56] the lifetime of Al wires is shown to be rather long up to 0.8 V, and then to decrease with increasing the bias. Since in our simulations we do not consider local current-induced heating, the calculated break-voltage of the wire corresponds to the voltage at which the energy barrier for the breaking process becomes zero, which corresponds to a vanishing life time in experiments. By extrapolating the experimental data for the lifetime as function of bias, shown in Fig. 7 of Ref. [56], one might estimate the experimental lifetime to vanish at around 1.6 V. A further detailed study of the breaking and of the electromigration phenomenon in Al nanowires with more realistic junction geometries, such as those discussed in Refs. [57, 58, 59], will be presented elsewhere.

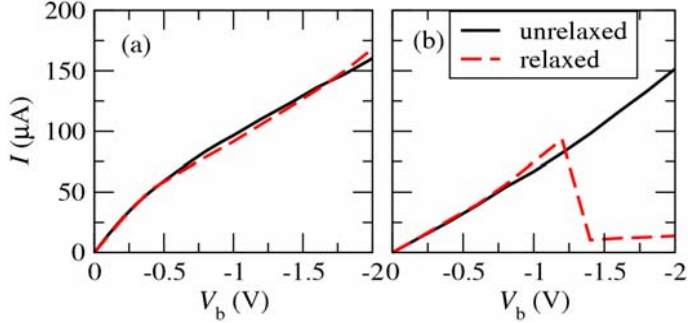

Fig. 8 (Color online) Current as function of the bias voltage, *I-V*, for the straight (a) and zigzag (b) configurations. The solid black line corresponds to calculations performed at fixed geometry while the red dashed curves are obtained by relaxing the geometry at each bias voltage.

Finally in Fig. 8 we discuss the effects produced by the structural relaxation on the electron transport properties of the wires. In particular we compare the *I-V* curves for both the straight and zigzag configurations obtained with a static geometry and by relaxing the structure under bias. For the straight chain the current is almost insensitive to structural relaxation since the atoms themselves move little. The situation is however different for the zigzag configuration. Interestingly the first structural transition at -0.2 V does not affect the current as both the geometries across the transition correspond to similar zigzag wires [compare Fig. 4(b) and Fig. 7(a)]. At -1.4 V however there is a drastic decrease of the current due to the wire breaking. If we take the van der Waals radius of Al, equal to 1.84 Å, as the radius of the Al monoatomic chain, we can estimate its cross section to be about 10 Å$^2$. The critical current density for the wire breaking is then calculated to be about $9.6 \times 10^{10}$ A/cm$^2$, in good agreement with the measured value[60] of $5 \times 10^{10}$ A/cm$^2$. Furthermore, at this point the transport changes from ballistic to tunnelling, with the residual bonding interaction being responsible for the non-zero current.

## V.  ELECTROMIGRATION OF A Si ATOM ON (4, 4) SWCNT

Carbon-based integrated circuits might be used as a complementary technological platform for silicon-based microelectronics.[61-63] For instance, carbon nanotube interconnects were successfully employed to bridge on-chip silicon transistors in realistic operational environments.[62,63] It is therefore desirable to explore the effects of current-induced forces over the electromigration of Si impurities on SWCNTs. In the last decade many theoretical studies have focused on calculating current-induced forces for single atomic small impurities (B, C, N, O and F) and alkali metal species sidewall adsorbed on CNTs.[22,23,64] To our knowledge however the possibility of electromigration has never been explored before. Here we present a series of calculations for the (4,4) metallic SWCNT incorporating a single silicon atom, sidewall adsorbed at the bridge position. By means of extensive optimization a user-defined double-ζ plus polarization basis set is constructed both for C and Si. We use an equivalent mesh cutoff of 400 Ry for the real space grid, while the cell simulation dimensions are set to be 20.0 Å, 20.0 Å and 19.846 Å respectively along the *x*, *y* and *z* directions (the transport direction is *z*). The initial relaxed atomic structure of a (4,4) SWCNT composed of 16 layers of carbon atoms [see Fig. 9] is obtained by conjugating gradient relaxation until any atomic forces is smaller than 0.03 eV/Å. The lead unit cell contains 4 layers with each layer comprising 8 C atoms (for this first calculation the leads are also (4,4) SWCNTs). Two stable independent adsorption sites at the bridge position are determined by structural relaxations at zero-bias. These are indicated as the A site ($Si_A$) and the B1 site ($Si_{B1}$) in Fig. 9. The difference between the two bonding sites is that the A site lies on a C-C bond perpendicular to the tube axis and therefore perpendicular to the transport direction, whereas the B1 site is on a bond slanted from the tube axis.

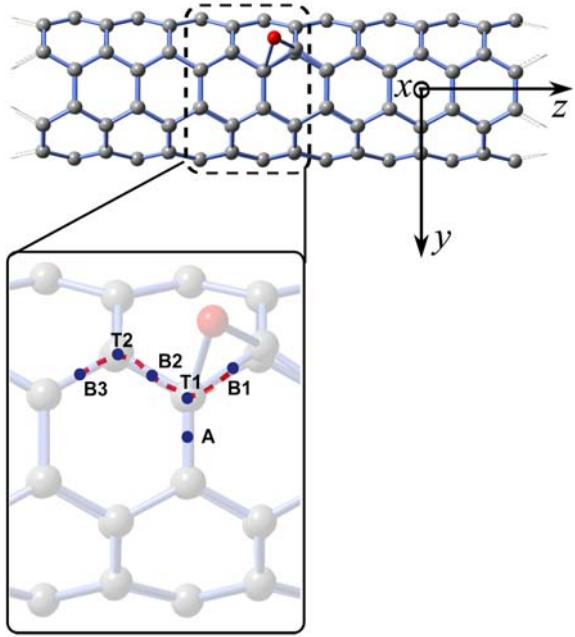

Fig. 9 (Color online) Schematic representation of an infinite (4,4) SWCNT with one Si atom sidewall attached at the B1 site. The zoom-in figure below shows the diffusion path of the Si atom from the B1 to B3 site (red dashed line). The adsorption site A is also represented.

For both the configurations we calculate the *I-V* curves up to a voltage of 1.5 V, while at the same time we relax the structure at each bias step. We find that $Si_{B1}$ starts to dramatically migrate along the SWCNT already at the rather low threshold bias of 0.5 V (with a current of about 60 μA). Since in our calculations we do not include the ionic vibrations caused by local heating, the migration of $Si_{B1}$ in our structural relaxation indicates a vanishing energy barrier along the migration path at 0.5 V. The migration path essentially involves positioning the Si atom alternatively at B1 positions and on top of C atoms [see path B1→B3 in Fig. 9]. In contrast, $Si_A$ remains almost still for all the bias voltages considered. An analysis of current-induced forces at finite bias indicates that the force acting on $Si_A$ along the C-C bond (the *y*-axis) is negligible and therefore the $Si_A$ atom does not move. In contrast for $Si_{B1}$ there is a substantial force along the C-C bond since it lies along the current flow. This causes migration at low bias.

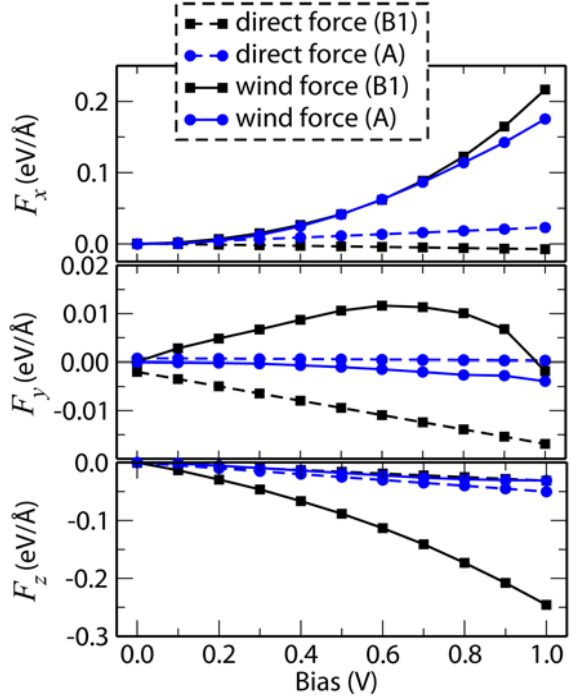

Fig. 10 (Color online) The current-induced forces acting upon the Si adatom as a function of bias are decomposed into electrostatic forces and wind forces along the three cartesian directions [see Eq. (16) and Eq. (17)].

In order to understand the nature of the current-induced forces acting on Si we decompose them into the two components described at the end of Sec. II, namely the electrostatic and the wind component (the total force is equal to the sum of the two). These are shown in Fig. 10 for both $Si_A$ and $Si_{B1}$ as function of bias. As expected the electrostatic force increases approximately linearly with bias, but it is rather small for both the two Si positions. This is due to the fact that the electric field is rather weak along the long CNT considered and also because the Si adatom is almost in a charge neutral state. Furthermore, in order to confirm the small contribution made by the electrostatic forces, we also calculate the electrostatic forces for both the $Si_A$ and $Si_{B1}$ adatoms by using Siesta, where an electric field is applied along the tube axis, for a CNT of finite length. The results show that the electrostatic forces in both cases are negligibly small. In contrast the wind force for the $Si_{B1}$ atom is large, so that in this case it almost coincides with the total force. We can then conclude that it is the wind force to be responsible for the electromigration of $Si_{B1}$.

Interestignly for $Si_A$ there is a significant wind force acting along the *x*-direction,

i.e. along the CNT radial direction. Si$_A$ is however tightly bound to the CNT along the radial direction and therefore it does not move. However, although the atom does change its position significantly away from the CNT, still a significant radial current-induced force should result in a measurable change in the desorption barrier height. This can be indeed measured in a STM experiment. We finally note that the *z*-component of the wind force acting on Si$_A$ is much smaller than that acting on Si$_{B1}$. Since the wind force depends on the current-induced electron charge redistribution around the given scattering center, one might expect that such redistribution will be small if there is a small current along a bond.

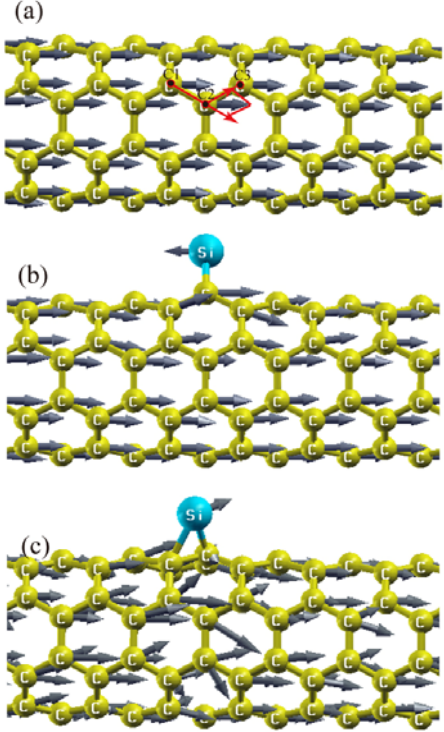

Fig. 11 (Color online) Schematic pictures of bond current distribution in three systems at the bias voltage of 1 V: (a) an infinite and perfect (4,4) SWCNT, (b) an infinite (4,4) SWCNT with Si$_A$ and (c) with Si$_{B1}$. The red lines in (a) illustrate that the total current vector on each C atom is mainly given by a sum of the bond-current vectors from its two longitudinally neighboring C atoms thus the transverse component in the total current vector vanishes.

In order to verify this hypothesis we calculate the bond current for the two configurations. The bond current between two orbitals, $J_{\mu\upsilon}$, is obtained as [65-67]

$$J_{\mu\upsilon} = \frac{2e}{h} \int_{-\infty}^{\infty} \left[ H_{\mu\upsilon} G^<_{M\upsilon\mu}(E) - H_{\upsilon\mu} G^<_{M\mu\upsilon}(E) \right] dE , \qquad (18)$$

while that between two atoms with indices *I* and *K*, $J_{IK}$, corresponds to the sum of the bond currents between individual orbitals located at those atoms

$$J_{IK} = \sum_{\mu \in I, \upsilon \in K} J_{\mu\upsilon} . \qquad (19)$$

Furthermore the total bond current acting on an individual atom can be represented in

vectorial form as

$$\vec{J}_I = \sum_{K \neq I} J_{IK} \vec{v}_{IK} , \qquad (20)$$

where $\vec{v}_{IK}$ is the vector connecting atoms *I* and *K*. The calculated bond currents for a bias voltage of 1 V are shown in Fig. 11. In the figure the dark arrows represent the current vector acting on an atom according to Eq. (20). First we show the bond currents for an infinite and defect-free (4,4) SWCNT [Fig. 11(a)] and find that the current flow is mainly along the longitudinal C-C bonds (the one along the SWCNT axis) with the transverse component vanishing. We then evaluate the bond currents for the SWCNTs with the Si adatoms. The presence of $Si_A$ perturbs only marginally the bond currents acting on the C atoms, which closely resemble those of the defect-free SWCNT [Fig. 11(b)]. This means that $Si_A$ acts as a weak scattering center, i.e. it is off the current flow. There are small bond currents between the Si atom and its neighboring C atoms, which results in a total bond current on $Si_A$ pointing against the overall current flow. Importantly, no currents pass across the transversal C-Si-C bonds either from C to Si or from C to C atoms, and this explains why the *y*-component of the force is very small. In contrast $Si_{B1}$ lies in the current flow path and therefore acts as a strong scattering center [Fig. 11(c)]. As a result the current on the SWCNT atoms significantly changes as compared to the case of the perfect SWCNT. This time there is a large bond current along the C-Si-C bonds and consequently a large *z*-component of the wind force.

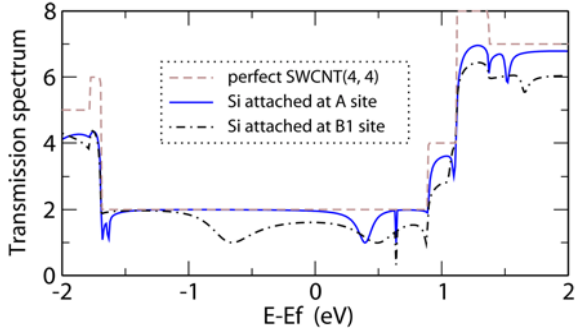

Fig. 12 (Color online) Zero-bias transmission coefficients as function of energy for the three systems illustrated in Fig. 11: an infinite and perfect (4,4) SWCNT, an infinite (4,4) SWCNT with either $Si_A$ or $Si_{B1}$.

Further support to our analysis of the scattering properties of the two Si adsorption centers is provided by comparing the difference between the corresponding zero-bias transmission coefficients [see Fig. 12]. We note that for $Si_{B1}$ there is a reduction in transmission around the Fermi energy significantly larger than that produced by $Si_A$. This is fully consistent with the previous finding that $Si_{B1}$ is a stronger scattering center than $Si_A$.

So far we have performed calculations for an infinite metallic (4,4) SWCNT with a single Si scatterer (this means that both the scattering region and the leads are formed by the same SWCNT). Clearly this does not correspond to a completely physical situation at finite bias as in our setup the potential imposed over the scattering region cannot be screened completely at the boundary with the leads.[68] Therefore, in order to verify whether or not the results presented in this section depend on these (artificial) boundary conditions, we set up a separate calculation, where the SWCNT is now attached to two Au electrodes. In this case the finite bias potential is fully screened by Au. The unit cell used for the $Si_{B1}$ case is shown in Fig. 13 and it is identical to the one used for $Si_A$. Importantly the results obtained for the infinite SWCNT are fully preserved, namely $Si_A$ does not move at any bias, whereas $Si_{B1}$ will migrate if the applied voltage is sufficiently large. The critical bias for the migration is now 0.8 V. This result is expected as the addition of two CNT/Au interfaces introduces a supplementary contact resistance (the current at the critical voltage for migration is about 45 μA). Most importantly the distribution of the bond currents around the Si adatom at 0.8 V is very similar to that at 0.5 V for the infinite SWCNT case, indicating once again that all the action is due to the large wind force.

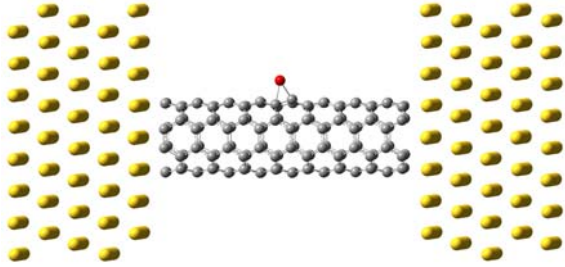

Fig. 13 (Color online) Two-terminal device constructed by attaching a SWCNT to two Au electrodes. The SWCNT includes also a $Si_{B1}$. This is equivalent to the device geometry of Fig. 9, where now the SWCNT leads are replaced by Au.

Finally, before concluding this section we analyze the effects of the Si adatom diffusion on the transport properties. We perform a structural relaxation for $Si_{B1}$ at $V_b$=0.5 V, and calculate the current at each conjugate gradient step. The resulting simulated steady-state current behaves in an oscillation form, reflecting the diffusion of the $Si_{B1}$ atom along the B1 to B3 path [see Fig. 9]. Our calculation reveals that the five turning points of the simulated current correspond to the specific atomic positions indicated in Fig. 14(a). The transient current for adsorption at the bridge site is about 50% larger than that for the top site. Finally we calculate the energy dependent transmission coefficients at 0.5 V for Si positioned on two selected locations along the migration path, namely the B2 and T1 sites [see Fig. 9]. The results are shown in Fig. 14(b). The transmission for adsorption at T1 is significantly reduced within the energy window between -0.5 eV and 0.5 eV when compared to that for B2 adsorption. This indicates that one of the two conducting channels incident from the nanotube lead is likely switched off near the Fermi level due to a localized impurity state around the silicon adatom.[69] From a scattering point of view, the Si adatom at the T1 position appears to be a much stronger scattering center for the current, and therefore at the considered voltage of 0.5 V it induces a larger force. This is 0.24 eV/Å, compared to the one for B2 adsorption, which is equal to only 0.10 eV/Å.

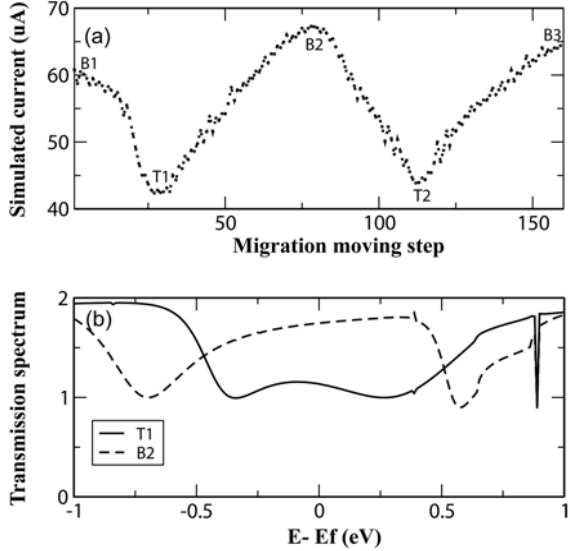

Fig. 14 (a) The current at $V$=0.5 V oscillates with the $Si_{B1}$ migration process; the characteristic transient Si bonding locations are seen as turning points; (b) transmission coefficients at $V$=0.5 V for Si adsorbed at the T1 and B2 position respectively [see Fig. 9].

## VI. CONCLUSIONS

In conclusion we have presented an algorithm for evaluating current-induced forces in atomic junctions and a few applications for systems of scientific and technological interest. The algorithm naturally integrates into the NEGF plus DFT framework and it is implemented in the *Smeagol* code. This enables us to perform atomic relaxations out of equilibrium in the presence of an electrical current and thereby to investigate the interplay between structural relaxation and transport properties. The algorithm is first thoroughly tested against known results and benchmarked against total energy calculations, whenever possible.

We have then taken on two systems of significant scientific interest. Firstly we have studied current-induced forces in Al nanowires either with a straight or a zigzag configuration and discussed their bias-induced structural instabilities. Importantly we have estimated a critical current density for the junction breaking rather close to the one measured experimentally. Finally we have explored the possibility for current-induced forces to manipulate the position of a Si adatom on the surface of a (4,4) metallic SWCNT. Remarkably our calculations predict electromigration as soon as the bias voltage exceeds a certain critical value. We have then demonstrated that the position dependent wind force is the one responsible for the diffusion process. A close analysis of the transmission has revealed that the wind-type current-induced forces are closely related to the local electron scattering strength.

**Acknowledgements**

The authors thank Tchavdar Todorov for useful discussions and for a critical reading of our manuscript. This project was supported by the National Natural Science Foundation of China (no. 61071012), the Ministry of Education (NCET-07-0014), the MOST of China (nos. 2007CB936204 and 2011CB933001) and the CSC program of China. SS and IR acknowledge the Science Foundation Ireland (Grant No. 07/IN.1/I945). Computational resources have been provided by the HEA IITAC project managed by the Trinity Center for High Performance Computing and by ICHEC.